\documentclass[aps,a4paper,10pt,twocolumn,nofootinbib]{revtex4} 
\usepackage{amsfonts}
\usepackage{amssymb}
\usepackage{mathrsfs}
\usepackage[mathscr]{euscript}
\usepackage[dvips]{graphicx}
\usepackage{fancyhdr}
\usepackage{makeidx}

\def\convol{\star}

\begin{document}

\title{Pulse propagation methods in nonlinear optics}
\author{P. Kinsler}
\email{Dr.Paul.Kinsler@physics.org}
\affiliation{
  Blackett Laboratory, Imperial College London,
  Prince Consort Road,
  London SW7 2AZ, 
  United Kingdom.}

\begin{abstract}

I present an overview of pulse propagation methods
 used in nonlinear optics, 
 covering both full-field 
 and envelope-and-carrier methods.
Both wideband and narrowband cases are discussed.
Three basic forms are considered -- 
 those based on (a) Maxwell's equations, 
 (b) directional fields, 
 and (c) the second order wave equation.
While Maxwell's equations simulators
 are the most general, 
 directional field methods can give significant 
 computational and conceptual advantages.
Factorizations of the second order wave equation 
 complete the set by being the simplest to understand.
One important conclusion is that 
 that envelope methods based on forward-only 
 directional field propagation has made the 
 traditional envelope methods (such as the SVEA, and extensions) 
 based on the second order wave equation utterly redundant.

\end{abstract}


\newcommand{\sech}{{\textrm{ sech}}}

\lhead{\includegraphics[height=5mm,angle=0]{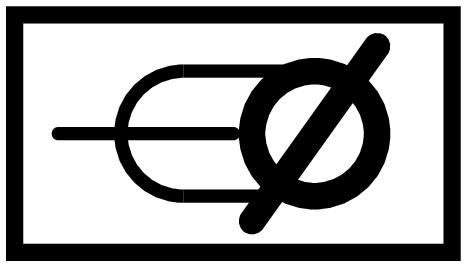}~~ENVEL}
\chead{~}
\rhead{
\href{mailto:Dr.Paul.Kinsler@physics.org}{Dr.Paul.Kinsler@physics.org}\\
\href{http://www.kinsler.org/physics/}{http://www.kinsler.org/physics/}
}
\lfoot{ }
\rfoot{Kinsler-2007-envel}

\date{\today}
\maketitle
\thispagestyle{fancy}

\tableofcontents

\chead{Pulse propagation methods in NLO}

%
\section{Introduction}

Here I discuss three important ways to tackle 
 pulse propagation in nonlinear optics.
These include methods that both do and do not 
 follow the traditional approach of using pulse envelopes.
The description is taken in the 1D limit, 
 but some discussion on including transverse effects is made.
The aim is to cover the considerations relevant when modeling 
 wideband fields, 
 a regime not comprehensively treated in many 
 standard texts
 \cite{Agrawal-NFO,Shen-PNLO,Boyd-NLO,Yariv-QE,Haus-WFOE,Siegman-Lasers}.

The three ways are solving (a) Maxwell's equations, 
 (b) directional Maxwell's equations, 
 or the (c) standard second order wave equation.
Solving Maxwell's equations is a well established approach, 
 with a long history 
 (i.e. finite difference time domain (FDTD), 
 see e.g. \cite{Gilles-HV-2000jcp}),
 although it is computationally intensive and has generally been little
 used in nonlinear optics 
 (but see e.g. \cite{Flesch-PM-1996prl,Gilles-MV-1999pre,Tyrrell-KN-2005jmo}).
Practical versions of directional Maxwell's equations have 
 appeared only recently, 
 such as that of 
 Kolesik et al. \cite{Kolesik-MM-2002prl,Kolesik-M-2004pre}; 
 other approaches followed
 \cite{Kinsler-RN-2005pra,Kinsler-2006arXiv-fleck,Mizuta-NOY-2005pra}, 
 the most general being \cite{Kinsler-2010pra-dblnlGpm}.
However the first proposal dates back to Fleck in 1970 \cite{Fleck-1970prb}, 
 although only as something of a remark in passing,
 rather than a full investigation. 
The most common approaches  nonlinear optics are those based on 
 the standard second order wave equation,
 particularly with regard to envelope propagation and the 
 celebrated slowly varying envelope approximation (SVEA).
The SVEA allows us to convert the second order wave equation
 into a first order equation that can efficiently
 propagate narrowband pulses.
Recently the SVEA has been 
 relaxed \cite{Brabec-K-1997prl,Porras-1999pra,Kinsler-N-2003pra}, 
 extending the use 
 to moderate bandwidths.
However, 
 much better approaches based on factorizing the second order 
 wave equation also exist.
An early example can be seen in \cite{Shen-PNLO}, 
 but also most notably by Blow and Wood \cite{Blow-W-1989jqe}, 
 and also the recent Ferrando et al. \cite{Ferrando-ZCBM-2005pre}
 and Genty et al. \cite{Genty-KKD-2007oe}; 
 the most general formulation, 
 even allowing for magnetic effects, 
 is that of Kinsler \cite{Kinsler-2010pra-fchhg}.

We can try to solve any of these equations directly, 
 without recourse to an envelope and carrier representation.
This means ensuring sufficient numerical resolution
 to integrate each of the field oscillations 
 as it propagates across the simulation window.
This approach is the standard one when solving 
 Maxwell's equations (i.e. FDTD), 
 but generally in nonlinear optics an envelope approach 
 is used.
This has a number of advantages: 
 it imposes a direction on the modeled pulse, 
 and it removes the fast oscillations at the centre frequency.
In combination with a moving frame, 
 it can turn a pulse of rapidly oscillating fields 
 moving at the speed of light 
 into a smooth, 
 nearly-stationary waveform -- 
 with commensurate gains in simulation speed.
These benefits usually come with a 
 restriction on the allowed bandwidth of the pulse 
 being modeled.

This paper is organized as follows: 
 in section \ref{S-envelopes} I compare
 field and envelope approaches.
In section \ref{S-maxwell} I consider Maxwell's equations
 in both field and envelope pictures, 
 followed in section \ref{S-directional} by the same, 
 but utilizing a directional rewriting of  Maxwell's equations.
Then, 
 in section \ref{S-secondorder} 
 I consider the role of second order wave equations, 
 in particular using factorization methods.
Finally, 
 in section \ref{S-conclusion} I present some conclusions.

Although not directly relevant to the discussion here, 
 it is also worth noting that directional waves in unstable resonators 
 were quantized by Brown and Dalton\cite{Brown-D-2002jmo}.

%
\section{Fields vs Envelopes}\label{S-envelopes}

It is often remarked upon that envelope methods work surprisingly well.
However, 
 this surprise seems to be based largely upon the SVEA equation 
 for pulse propagation, 
 which indeed contains many approximations 
 (see \cite{Kinsler-2002-fcpp,Kinsler-N-2003pra}).
Recently it has been shown by several groups 
 \cite{Kolesik-MM-2002prl,Ferrando-ZCBM-2005pre,Kinsler-RN-2005pra,Mizuta-NOY-2005pra} 
 that equations nearly identical to those generated by the 
 SVEA can be found by assuming little more than the 
 lack of backward-going field components.
Even when revisiting Blow and Wood \cite{Blow-W-1989jqe},
 we can see that their mathematics contained minimal constraints on the 
 bandwidth of the envelope, 
 although their specific nonlinearity model did contain such restrictions.

%
\subsection{The definition }\label{Ss-envelopes-def}

For envelope methods,
 the direction is imposed by the form for the carrier function, 
 and is usually a plane wave traveling in the chosen direction. 
Thus the typical envelope and carrier representation of some field $Q$ 
 is
~
\begin{eqnarray}
  Q(t; z)
&=&
  A(t; z)
  e^{\imath\left(k_0 z-\omega_0 t\right)} 
 +
  A^*(t;z)
  e^{-\imath\left(k_0 z-\omega_0 t\right)} 
,
\label{eqn-Q-env0}
\\
  \tilde{Q}(\omega;z)
&=&
  \tilde{A}(\omega+\omega_0; z) e^{\imath k_0 z } 
 +
  \tilde{A}^*(\omega-\omega_0; z) e^{-\imath k_0 z } 
.
\label{eqn-Q-env}
\end{eqnarray}
In some of the following equations, 
 I will shorten the argument of the exponential in the 
 carrier function with $\Xi = \imath\left(k_0 z-\omega_0 t\right)$.
It is worth noting that we are not required to use carrier functions
 with the usual exponential form \cite{Gabor-1946jiee}, 
 e.g. in semiconductor physics,
 Bloch functions are routinely used 
 as carriers to form a basis for electron (or hole) envelope functions.

Note that it is {\em approximations} that restrict the validity of 
 envelope approaches, 
 not the use of them in itself.
This is contrary 
 to the impression that might be gained from 
 SVEA approaches, 
 and even their generalizations \cite{Brabec-K-1997prl,Kinsler-N-2003pra}.
The potential benefits of envelopes are not 
 tied to restrictions
 on the bandwidth of the pulse being modeled.

%
\subsection{The big advantage}\label{Ss-envelopes-advantange}

Replacing real fields $E$ and $H$ with an envelope-carrier 
 description give us at least one clear advantage: 
 it removes the dominant contribution to the underlying field oscillations.
The resulting smoother envelope is therefore easier, 
 and much less computationally expensive to propagate.
It is the rapidity of the fastest time-domain modulation of the 
 field or envelope which constrains our time resolution, 
 and the rapidity of the fastest spatial modulation which 
 constrains the spatial resolution.
Note that 
 although we usually hope that our envelope will then 
 have a relatively slowly varying form,
 the mere replacement of the EM fields by envelope-carrier combinations
 imposes of itself no approximations whatsoever.

Two processes may act to twist an initially smooth envelope
 into something more problematic.
First, 
 linear dispersion can add chirp,
 which manifests itself as a nonlinear phase ramp across the pulse.
These are usually relatively smooth changes, 
 and cause little problem.
Second, 
 there are nonlinear effects.
Some, 
 such as self phase modulation (SPM) can be relatively mild, 
 others, 
 such as coupling to backward propagating waves and harmonic generation
 can impose significant oscillations.

%
\subsection{Nonlinear polarization terms}\label{Ss-envelopes-nonlniearity}

As mentioned above, 
 nonlinear processes can generate oscillatory contributions 
 to the envelope.
We can see how this occurs by considering an instantaneous 
 third order nonlinearity, 
 which depend on $E^3$ and has the form
~
\begin{eqnarray}
  E^3(t; z)
&=&
 \left[
  A(t; z)
  e^{\Xi}
 +
  A^*(t;z)
  e^{-\Xi}
 \right]^3
\label{eqn-envelopes-nonlinear3-a}
\\
&=&
  A(t; z)^3
  e^{+3\Xi}
 +
  3
  A(t; z)^2
  A^*(t;z)
  e^{+\Xi}
 +
  \text{c.c}
\label{eqn-envelopes-nonlinear3-b}
.
\end{eqnarray}
Here we see 
 a useful side-effect of the envelope-carrier representation --
 that nonlinear terms can be separated into convenient components.
In this example, 
 the full $\chi^{(3)}$ nonlinearity splits into 
 a third harmonic generation (THG) term proportional to $A^3$, 
 and an SPM term proportional to $\left|A\right|^2 A$; 
 along with complex conjugate counterparts (c.c).
Clearly the THG term is non-resonant with the chosen carrier, 
 and keeping such non-resonant nonlinear terms will impose 
 significant oscillations onto our envelope as it propagates. 
Such oscillations break approximations 
 relying on a relatively smooth envelope, 
 which is why in SVEA models they are discarded; 
 however there is no {\em a priori} requirement to do so.
E.g., 
 in the wideband Raman model 
 of Kinsler and New \cite{Kinsler-N-2005pra,Kinsler-2005eWBRAM},
 the Stokes and anti-Stokes fields appeared as sidebands on the 
 envelope spectrum.

%
\subsection{Multiple carriers, wideband envelopes}\label{Ss-envelopes-multi}

We can generalize from a single envelope-carrier pair 
 by using multiple envelopes with carriers at different carrier frequencies.
However, 
 each added envelope greatly increases the number of 
 individual polarization terms resulting from a nonlinearity, 
 so as a rule it is best to 
 use the minimum number possible.
As an exercise, 
 just calculate the number of terms in an 
 eqn. (\ref{eqn-envelopes-nonlinear3-a})
 derived from a field defined as 
 $E = A_1 e^{+\Xi_1} + A_2 e^{+\Xi_2} + \text{c.c}$,  
 not $E = A_1 e^{+\Xi_1} + \text{c.c}$!

Multiple carriers work best when there are 
 multiple narrowband fields which 
 resonantly interact, 
 such as in an optical parametric amplifier 
 \cite{Shen-PNLO,Boyd-NLO} (check)
 or for Raman processes \cite{Kinsler-N-2005pra,Kinsler-2005eWBRAM}.
In such cases we can ruthlessly discard nonlinear polarization terms 
 which are not perfectly in resonance with processes of our choosing.
Another use for multiple carriers is for including both 
 forward and backward propagating fields.

If we use multiple carriers, 
 and also allow wideband envelopes, 
 then it is possible for multiple envelopes to cover the same
 piece of spectrum.
In a continuous mathematical description this overlap will always happen, 
 but in a discrete or numerical implementation it will depend on 
 our parameters.

This overlap is not necessarily a problem, 
 as long as we are careful about assigning polarization terms 
 to whichever envelope equation we choose --
 we must make sure not to add the same term twice, 
 for example.
This can also lead to a non-unique description,
 in the case where a polarization term could be equally well drive 
 the evolution of one of two (or more) envelopes.
Nevertheless, 
 such non-uniqueness does not break our model, 
 it just allows us a choice which we might be able to use
 to our advantage.
In a simulation, 
 we could try to assist our numerics
 by managing the envelope spectra by 
 generating a total spectrum, 
 and then reassigning components in the overlap region according
 to some smoothing procedure.

The use of wideband envelopes can raise some interesting issues.
For example, 
 if the bandwidth of an envelope is greater than its carrier frequency, 
 then the envelope will extend into negative frequencies.
This is not a problem for our physical model, 
 since we still have to reconstruct the field from the envelopes and carriers, 
 and those negative frequency components are matched by complementary 
 positive frequency ones
 from the complex conjugate of the envelope\footnote{That is, 
   they {\em should} be so matched.  
  If they aren't, you've done something wrong.}.
Again, 
 we might consider a spectral management scheme which swaps these 
 unexpected components over, 
 restoring the envelope to pure positive frequency content
 (and hence its conjugate to pure negative frequency content).
However, 
 we then find that at zero-frequency we have introduce a hard cutoff
 in the envelope spectra, 
 and so induced unwanted oscillations
 in the time domain version of the envelope.

However, 
 while such spectral management might seem to offer advantages, 
 in practice it makes little difference, 
 and adds needless complication to simulation code.
I would consider it only if some unexpected interaction was 
 generating significant spectral content near the band edge of an envelope, 
 at a position where it would be well within the spectral range of some 
 other envelope; 
 and even then it might be easier 
 to simply increase the envelope bandwidth.

%
\subsection{Directionality}\label{Ss-envelopes-directionality}

A carrier imposes a direction of propagation, 
 and most carrier-based models silently neglect even the possibility 
 of backward propagating fields, 
 even though there is a coupling between them.
However, 
 Casperson\cite{Casperson-1991pra} used both forward and backward carriers
 to construct an envelope model with a separation of the 
 forward and backward field components and interactions.
The more recent paper of Sanborn et al. \cite{Sanborn-HD-2003josab}
 used the same approach.

Backward traveling components, 
 if forced onto a forward traveling envelope, 
 will appear as non-resonant terms.
If identified correctly, 
 these can then be discarded.

See also section \ref{S-forwardbackward} for a discussion of 
 the coupling between forward and backward waves which is 
 induced by a nonlinearity.

%
\subsection{Moving frames}\label{Ss-envelopes-moving}

In combination with a suitable moving frame, 
 an envelope representation can
 turn a pulse of rapidly oscillating fields 
 moving at the speed of light 
 into a smooth, 
 nearly-stationary waveform -- 
 with commensurate gains in simulation speed.

However, 
 we need to guarantee that all contributions from backward
 traveling components are removed, 
 otherwise the envelope will contain 
 oscillatory components moving at approximately twice 
 the frame speed.

A typical moving frame is defined by for a frame speed $v$
 as $t' = t - z / v $.
Thus the spatial derivatives in propagation equations 
 are altered using
~
\begin{eqnarray}
  \partial_{z} 
&=&
  \partial_{z'} - v \partial_t
\end{eqnarray}

%
\subsection{Estimating the computational cost}\label{Ss-envelopes-compute}

Consider a wideband pulse, 
 with a bandwidth of the order of its centre frequency $\omega_0$.
In a full-field approach, 
 this will have the fastest modulations of the field
 being of the order $2\omega_0$.
In comparison, 
 an envelope approach results in 
 the fastest modulations on the envelope 
 being of the order $\omega_0$.
Simplistically we might then hope that the 
 envelope approach allows us to halve our time and space 
 resolutions whilst still retaining numerical accuracy.
For narrow band fields the advantage is much clearer --
 a bandwidth of $\omega_0/100$ might allow resolutions to be
 coarsened by a factor of 100.
For fields of a wider bandwidth, 
 we gain little advantage, 
 unless we shift our carrier frequency $\omega_0$ to the 
 centre of the spectrum, 
 even if that is not co-incident with the dominant 
 frequency component.
A more comprehensive examination of the effects 
 of numerical resolution has been given for 
 the nonlinear Schr\"odinger equation by 
 Sinkin et al. \cite{Sinkin-HZM-2003jlt}.

Note that since the linear response (dispersion) of the medium can 
 be done exactly in the frequency domain, 
 regardless of step-size, 
 it might seem more appropriate to focus more on 
 the role of the nonlinear response when 
 estimating the necessary temporal and spatial resolutions.
However the accuracy of a propagation method
 cannot be easily evaluated whilst ignoring 
 the dispersive propagation, 
 because both dispersive and nonlinear effects 
 occur simultaneously.
Even if a split step method is used, 
 they are interleaved, 
 and their effects cannot be disentangled.

%
\subsection{Disadvantages}\label{Ss-envelopes-disadvantage}

The slight disadvantage of using envelopes is that real valued 
 time dependent fields
 are replaced by complex valued envelopes.
This doubles the amount of storage used during computations, 
 and also requires the use of complex Fourier transforms
 rather than the faster real ones; 
 although of course the spectra of the fields are complex
 in any case.
In practice, 
 the computational cost is small, 
 although the complexity of the simulation code is increased.

%
\section{Maxwell's equations}\label{S-maxwell}

When propagating  fields in free space, 
 we use the the source-free Maxwell's equations.
To simplify the description we transform their time-like
 behaviour into frequency space.
This enables us to write the convolutions required to 
 model the linear time-response of the medium (e.g. dispersion) 
 as multiplications.
However, 
 since the form of the nonlinear response is not simplified by this process, 
 a convolution in frequency space appears. 
In frequency space, 
 time derivatives convert to factors of $-\imath \omega$,
 so the equations are
~
\begin{eqnarray}
  \partial_z \tilde{H}_y(\omega;z)
&=&
  - \imath \omega 
    \tilde{\epsilon}(\omega') 
    \convol 
    \tilde{E}_x(\omega;z)
,
\label{eqn-firstorder-dtE}
\\
  \partial_z \tilde{E}_x(\omega)
&=&
  - \imath \omega  
    \tilde{\mu}(\omega')  
    \convol 
    \tilde{H}_y(\omega;z)
.
\label{eqn-firstorder-dtH}
\end{eqnarray}
The ``$\convol$'' denotes a convolution,
~
\begin{eqnarray}
   Q(\tau)
  \convol 
   P(t)
&=&
  \int
    Q(\tau) P(t-\tau) d\tau
~
=
  \mathscr{F}^{-1}
  \left[
    \tilde{Q}(\omega) 
    \tilde{P}(\omega) 
  \right]
. ~~~~
\end{eqnarray}

A rather nice way to scale these equations is to define 
 suitable $\epsilon_n$ and $\mu_n$ corresponding to a suitably chosen 
 refractive index, 
 hence $\mu_n$ will usually be $\mu_0$. 
This means $c_n = 1 / \epsilon_n \mu_n$,
 $k = \omega / c_n$,
 $\tilde{\epsilon}_n = \tilde{\epsilon}(\omega')/\epsilon_n$
 $\tilde{\mu}_n = \tilde{\mu}(\omega')/\mu_n$.
We then define $e = \sqrt{\epsilon_n} E$
 and $h = \sqrt{\mu_n} H$, 
 which ensures $e$ and $h$ are of comparable sizes.
This gives us the scaled Maxwell's equations
~
\begin{eqnarray}
  \partial_z \tilde{h}_y(\omega;z)
&=&
  - \imath 
    k
    \tilde{\epsilon}_n(\omega') 
    \convol 
    \tilde{e}_x(\omega;z)
,
\label{eqn-firstorder-dtEE}
\\
  \partial_z \tilde{e}_x(\omega)
&=&
  - \imath 
    k
    \tilde{\mu}_n(\omega')  
    \convol 
    \tilde{h}_y(\omega;z)
.
\label{eqn-firstorder-dtHH}
\end{eqnarray}
It is worthwhile comparing this scaling with that from the 
 directional fields approach in section \ref{S-directional}; 
 with the correspondences $\sqrt{\epsilon_n} \leftrightarrow \alpha_r$
 and $\sqrt{\mu_n} \leftrightarrow \beta_r$.


For our purposes,
 there are two main ways to solve Maxwell's equations: 
 either FDTD\cite{Gilles-HV-2000jcp} or 
 Pseudo-Spectral Spatial Domain (PSSD)\cite{Tyrrell-KN-2005jmo}.
In FDTD we propagate forward in time, 
 holding the fields $E(z), H(z)$ as a function of space.
However, 
 in nonlinear optics, 
 it is more convenient to use PSSD, 
 where we propagate forward in space,
 holding the fields $E(t), H(t)$ as a function of time.
Under PSSD 
 derivatives are calculated pseudospectrally \cite{Fornberg-PSmethods}.
However, 
 its most important feature is that 
 the entire time-history (and therefore frequency content) 
 of the pulse is known at any point in space, 
 so applying even arbitrary dispersion incurs no extra computational penalty.
In contrast,
 FDTD (or other temporally propagated methods) must use convolutions 
 or time-response models for dispersion.
Although spatially propagated simulations (e.g. PSSD) 
 make it difficult to incorporate reflections properly, 
 this is not a significant constraint as most such simulations 
 are only interested in uni-directional propagation anyway.

For example, 
 in a 1D medium with linear dispersive properties 
 defined by $\epsilon_r$, $\mu_r$, 
 containing a third order $\chi^{(3)}$ nonlinearity
 defined by $\epsilon_c$, 
 the equations are
~
\begin{eqnarray}
  \partial_z 
    {H}_y(\omega;z)
&=&
 -
  \imath \omega \tilde{\epsilon}_r(\omega') \tilde{E}_x(\omega;z)
\nonumber
\\
&&
 -
  \imath \omega 
  \left\{
    \tilde{\epsilon}_c(\omega') 
    .
    \mathscr{F}
    \left[
      E_x(t;z)^2
    \right]
    (\omega) 
  \right\}
  \convol
  \tilde{E}_x(\omega;z)
,
 ~~~~ 
\label{eqn-pssd-dH}
\\
  \partial_z 
    \tilde{E}_x(\omega;z)
&=&
 -
  \imath \omega \tilde{\mu}_r(\omega') \tilde{H}_y(\omega;z)
,
\label{eqn-pssd-dE}
\end{eqnarray}
 where $\mathscr{F}[Q(t)](\omega)$ denotes the Fourier transform 
 of some function $Q(t)$.
This model allows for the time-response of 
 the nonlinearity, 
 and is thus applicable to (weakly coupled) Raman systems as well.
Note that the terms dependent on $\epsilon_r$ and $\mu_r$ 
 are simple products.
The linear dispersion combined with a time-dependent 
 third order nonlinearity gives a permittivity function which would be written
~
\begin{eqnarray}
  \epsilon(\tau, t)
&=&
  \epsilon_r(\tau) + \epsilon_c(\tau) \convol E(t)^2
\label{eqn-epsilon-t}
\\
  \epsilon(\tau, t) \convol E(t)
&=&
  \epsilon_r(\tau) \convol E(t)
 +
  \left\{
    \epsilon_c(\tau) \convol E^2 (t)
  \right\}
  E(t)
\label{eqn-epsilon-t-E-t}
\\
  \tilde{\epsilon}(\omega') 
  \convol
  \tilde{E}(\omega)
&=&
  \tilde{\epsilon}_r(\omega) \tilde{E}(\omega)
\nonumber
\\
&&
~~
 +
  \left\{
    \tilde{\epsilon}_c(\omega') 
   \mathscr{F}
     \left[
       E^2 (t)
     \right]
     (\omega') 
  \right\}
 \convol
  \tilde{E}(\omega)
\label{eqn-epsilon-w-E-w}
.
\end{eqnarray}
In the case of instantaneous nonlinearity, 
 $\tilde{\epsilon} \tilde{E} 
  = \tilde{\epsilon}_r \tilde{E} + \epsilon_c \mathscr{F}
    \left[E^3\right]$.

A simple and efficient way to propagate these equations is using 
 staggered $E$ and $H$ fields, 
 which allow us to use an Euler-like integration for each field, 
 but achieves second-order accuracy \cite{Yee-1966tap}.
However, 
 while the $E$ and $H$ fields necessary for a forward propagating
 pulse are easy to determine for co-incident $E$ and $H$, 
 we need to use staggered initial conditions or else 
 we get a significant backward propagating component.
Even with correctly staggered initial conditions, 
 we see a small spurious backward component, 
 the size of which depends on the time step.
This backward pulse is hard to get rid of completely, 
 but it can be filtered in the time domain when the 
 two pulses have propagated apart far enough.
Another point to consider, 
 particularly when generating the initial conditions
 for very short pulses, 
 is the zero-force condition \cite{ZFC}.
This can be easily satisfied by deriving the $E$ and $H$ fields for the 
 pulse from a suitable vector potential,  
 rather than simply assuming a form for the $E$ field.

When considering the solution of these these Maxwell's equations, 
 it is useful to partly calculate 
 the time derivative of eqn.(\ref{eqn-epsilon-t-E-t}).
For dispersion and a time response $\chi^{(3)}$
 this gives us three terms,
~
\begin{eqnarray}
 \partial_t
 \left(
  \epsilon(\tau, t) \convol E(t)
 \right)
&=&
 \partial_t
  \epsilon_r(\tau) \convol E(t)
\nonumber
\\
&&
~~
 +
 \left(
  \partial_t
  \left\{
    \epsilon_c(\tau) \convol E^2 (t)
  \right\}
 \right)
  E(t)
\nonumber
\\
&&
~~ ~~
 +
  \left\{
    \epsilon_c(\tau) \convol E^2 (t)
  \right\}
 \left(
  \partial_t
  E(t)
 \right)
.
\label{eqn-dt-epsilon-t-E-t}
\end{eqnarray}
Thus we see that to solve the equations, 
 we will need to calculate the derivatives of three terms:
 the usual dispersive term, 
 the time response term, 
 and the field.
We will also need to retain the value of the time response term as well.
Since the time response term contains a convolution, 
 it is best calculated in the frequency domain, 
 which is particularly convenient when using pseudospectral derivatives.
We will need {\em two} FFT's to transform $E$ and $E^2$ into frequency space.
 There we construct the dispersion term and the time response term 
 by simple multiplications, 
 and set up arrays for the derivatives by multiplying by $-\imath \omega$.
We then need {\em four} back transforms for a total of six in all:
 one more for the time response, 
 and three for the time derivatives of the 
 dispersion, time response, and field.
For an instantaneous nonlinearity, 
 we need only three FFT's:
 two forward transforms (for $E$ and $E^3$), 
 and one back transform for the combined 
 derivative.
In addition to these six (or three) FFT's needed to solve the 
 $\partial_z H$ equation, 
 the $\partial_z E$ equation requires another two, 
 for a total of eight (or five).

%
\subsection{Envelopes}\label{Ss-maxwell-envelopes}

Although it is not often done, 
 we can represent Maxwell's equations using an 
 envelope and carrier representation.
We express the fields $E$ and $H$
 using
~
\begin{eqnarray}
  E_x(t;z)
&=&
  A(t;z)
  e^{\imath\left(k_0 z-\omega_0 t\right)} 
 +
  A^*(t;z)
  e^{-\imath\left(k_0 z-\omega_0 t\right)} 
,
\label{eqn-E-env}
\\
  H_y(t;z)
&=&
  F(t;z)
  e^{\imath\left(k_0 z-\omega_0 t\right)} 
 +
  F^*(t;z)
  e^{-\imath\left(k_0 z-\omega_0 t\right)} 
.
\label{eqn-H-env}
\end{eqnarray}

We insert these into the Maxwell's equations above, 
 separate out the normal and complex conjugate (c.c.) parts,
 cancel the carrier exponentials present on both sides of the equations,
 and rearrange to leave only $\partial_z$ terms on the RHS,
~
\begin{eqnarray}
  \partial_z \tilde{F}(\omega;z)
&=&
 - 
  \imath \omega 
  \tilde{\epsilon} (\omega')
  \convol
  \tilde{A}(\omega;z)
 -
  \imath k_0
  \tilde{F}(\omega;z)
,
\label{eqn-envel-B}
\\
  \partial_z \tilde{A}(\omega;t)
&=&
 -
  \imath \omega 
  \tilde{\mu} (\omega') 
  \convol
  \tilde{F}(\omega;t)
 -
  \imath k_0
  \tilde{A}(\omega;t)
.
\label{eqn-envel-A}
\end{eqnarray}

Of course there is still much detail hidden in 
 the permittivity $\tilde{\epsilon}$, 
 since it contains the nonlinearity.
Consequently, 
 I do not apply this envelope definition to a general equation of motion
 because how $\epsilon$ is expressed usually depends on the field and
 therefore on those envelopes.
Starting with eqn.(\ref{eqn-epsilon-t-E-t}), 
 and expanding ${\epsilon}$ with terms
 for both (linear) dispersion $\epsilon_r$
 and a time dependent $\chi^{(3)}$ nonlinearity ($\epsilon_c$) gives
~
\begin{eqnarray}
   \epsilon(\tau, t)
  \convol 
   A(t) e^{+\Xi}
  &+&
  \textrm{c.c.}
\nonumber
\\
&=&
   \epsilon_r(\tau) 
  \convol 
   \left\{
     A(t) e^{+\Xi} 
   +
    A(t)^* e^{-\Xi}
   \right\}
\nonumber
\\
&&
 +
 \left(
     \epsilon_c(\tau)  
     \convol 
     \left\{
       A(t)^2 e^{+2\Xi} 
     \right.
 \right.
\nonumber
\\
&&
 \left.
     \left.
       + A(t) A(t)^* +  A(t)^{*2} e^{-2\Xi}
     \right\}
  \right)
\nonumber
\\
&& 
 \times
   \left\{
     A(t) e^{+\Xi} + A(t)^* e^{-\Xi}
   \right\}
\\
   \epsilon(\tau,t)
  \convol 
   A(t) 
   e^{+\Xi}
&=&
   \epsilon_r(\tau) 
  \convol
     A(t) e^{+\Xi}
\nonumber
\\
&& 
 +
   2
   \left\{
     \epsilon_c(\tau) 
    \convol 
      \left|A(t)\right|^2
   \right\}
     A(t) e^{+\Xi}
\nonumber
\\
&& 
 +
   \left\{
     \epsilon_c(\tau) 
    \convol 
      A(t)^2
   \right\}
     A(t)^* e^{+\Xi}
\nonumber
\\
&& 
 +
   \left\{
     \epsilon_c(\tau) 
    \convol 
      A(t)^2
   \right\}
     A(t) e^{+3\Xi}
\\
  \tilde{\epsilon}(\omega+\omega_0)
  \tilde{A}(\omega)
&=&
   \tilde{\epsilon}_r(\omega'+\omega_0)
  \convol 
   \tilde{A}(\omega)
\nonumber
\\
&& 
 +
   2
  \left\{
   \tilde{\epsilon}_c(\omega'+\omega_0) 
   \mathscr{F}
   \left[
      \left|A(t)\right|^2
   \right]
   (\omega')
  \right\}
  \convol
   \tilde{A}(\omega)
\nonumber
\\
&& 
 +
  \left\{
   \tilde{\epsilon}_c(\omega'+\omega_0) 
   \mathscr{F}
   \left[
      A(t)^2
   \right]
   (\omega')
  \right\}
  \convol
   \tilde{A}(\omega)^*
\nonumber
\\
&& 
 +
  \left\{
   \tilde{\epsilon}_c(\omega'+3\omega_0) 
   \mathscr{F}
   \left[
      A(t)^2
   \right]
   (\omega')
  \right\}
  \convol
   \tilde{A}(\omega)
.
\nonumber
\\
\label{eqn-epsilon-envelope}
\end{eqnarray}

We can see in eqn. (\ref{eqn-epsilon-envelope}) that 
 the first three of the terms (one dispersion and two SPM-like) 
 are resonant with the chosen envelope, 
 but the last (third harmonic generation) is not,
 and it modulates the envelope at $2\omega_0$, 
 (and subsequently the propagation by $\sim2k_0$).
Note the form of the second SPM-like term, 
 which needs contributions from two $A(t)$'s and one $A(t)^*$
 to have the correct frequency dependence, 
 but the convolution is with the $A(t)^*$ and not an $A(t)$
 as might be expected.

Note that in the case of instantaneous $\chi^{(3)}$, 
 the third RHS term reduces to $\epsilon_c \left|A(t)\right|^2 A(t)$,
 giving
~
\begin{eqnarray}
   \epsilon(\tau)
  \convol 
   A(t) e^{+\Xi}
&=&
   \epsilon_r(\tau) 
  \convol
     A(t) e^{+\Xi}
 +
   3
     \epsilon_c
     \left|A(t)\right|^2
     A(t) e^{+\Xi}
\nonumber
\\
&& 
 +
     \epsilon_c
      A(t)^3
      e^{+3\Xi}
,
\\
   \tilde{\epsilon}(\omega'+\omega_0)
  \convol
   \tilde{A}(\omega)
&=&
   \tilde{\epsilon}_r(\omega+\omega_0)
   \tilde{A}(\omega)
\nonumber
\\
&& 
 +
   3
     \epsilon_c
     \mathscr{F}
     \left[
       \left|A(t)\right|^2 A(t)
     \right]
     (\omega)
\nonumber
\\
&& 
 +
     \epsilon_c
     \mathscr{F}
     \left[
      A(t)^3
     \right]
     (\omega)
.
\end{eqnarray}

This expression can then be substituted directly into 
 eqns. (\ref{eqn-envel-B},\ref{eqn-envel-A})

In this formulation, 
 we have made no ``slowly varying'' approximation
 like those in traditional approaches
 \cite{Agrawal-NFO,Shen-PNLO,Boyd-NLO,Yariv-QE,Haus-WFOE,Siegman-Lasers}, 
 or in the variously corrected extensions\cite{Brabec-K-1997prl,Kinsler-N-2003pra}.
The price we pay is having two envelopes instead of one, 
 since now the magnetic field is explicitly retained.
Also, 
 the model still contains backward propagating components; 
 which, 
 with the chosen carrier functions, 
 will impress oscillations at $2 \omega_0$ on the envelope, 
 and oscillations of $2 k_0$ on the propagation.
 placing greater demands on our numerics.
Unfortunately there is no way to filter these out at 
 any point in the simulation,
 because their backward propagating nature can only be established
 by linking the time-like behaviour and space-like propagation
 of both $E$ and $H$ fields.
We cannot always rely on only time-like behaviour to filter them out, 
 because, e.g.,
 both backward propagating terms (at $+k_0$ and $-\omega_0$) 
 and third harmonic generation (at $+3k_0$ and $3\omega_0$) 
 are equally detuned from the carrier (at $+k_0$ and $\omega_0$);
 although we could do so if we were in a regime where 
 third harmonic generation were negligible.

At the start of this subsection, 
 we hoped that dividing out the carrier oscillations would give us a 
 slowly varying pulse envelope, 
 which would then enable us to coarsen our numerical resolution, 
 and speed simulations.
This is true, 
 up to a point -- 
 but remember the most likely reason we are using a Maxwell solver is 
 that we want to model a wideband situation.
It is the rapidity of the fastest time-domain modulation of the 
 field or envelope which constrains our time resolution, 
 and the rapidity of the fastest spatial modulation which 
 constrains the spatial resolution.

We can do better than these Maxwell equations approaches without 
 having to use second order wave equations 
 and their complicated approximations 
 by using directional Maxwell's equations, 
 as described in the next section.

%
\subsection{Transverse effects}
\label{S-maxwell-transverse}

There are two main transverse effect likely to be of interest 
 in pulse propagation models: 
 mode averaging, 
 and diffraction or off-axis propagation.

Mode averaging is easy to incorporate if 
 you assume some known transverse profile for the mode: 
 e.g. for an optical fibre or some other waveguide.
The transverse derivatives vanish, 
 and the material properties are evaluated as an 
 integral over the transverse dimensions, 
 weighted by the mode function.

Diffraction and off-axis propagation 
 they result from a coupling between the vector components of 
 the $E$ and $H$ fields -- 
 including those along the propagation direction.
Thus they are much harder to understand, 
 as compared to a paraxial model based on (e.g.) 
 the second order wave equation, 
 although they can be simulated easily enough in a full 4D 
 FDTD code.
This is because
 they result from a coupling between the vector components of 
 the $E$ and $H$ fields -- 
 including those along the propagation direction.

%
\section{Directional Maxwell's equations}\label{S-directional}

To my knowledge, 
 the earliest rewriting of Maxwell's equations in a directional form 
 was by Fleck \cite{Fleck-1970prb}, 
 who treated a dispersionless medium and plane polarized wave.
However, 
 the idea was not used beyond its brief appearance there.
Fleck constructed his directional fields by combining 
 the sum and difference of the E and H fields, 
 weighted by the square roots of the permittivity $\epsilon$ 
 and permeability $\mu$ respectively.
The new combined fields represent
 the forward and backward traveling components of the total field, 
 and we can derive 
 first-order wave equations for these new fields.

In the mid 1990's, 
 the concept was rediscovered and used to evaluate the properties of 
 grating structures by de Sterke, Sipe, and co-workers 
 \cite{Sipe-PD-1994josaa,deSterke-SS-1996pre}, 
 but not applied to pulse propagation.
The work considered materials with a spatially varying refractive index, 
 but did not incorporate material dispersion or nonlinearity.

This concept of using directional fields for {\em pulse propagation} 
 was not revisited until the work of 
 Kolesik et al. \cite{Kolesik-MM-2002prl,Kolesik-M-2004pre}. 
After selecting a preferred direction,
 they then projected out the forward-like and backward-like parts of 
 the propagating fields.
This procedure resulted in first order wave
 equations for the propagation of the forward and backward field components.
Subsequent work by 
 Kinsler et al. \cite{Kinsler-RN-2005pra,Kinsler-2006arXiv-fleck}, 
 presented a directional rewriting
 of Maxwell's equations
 using a generalized form of Fleck's construction; 
 note also the independent work of 
 Mizuta et al. \cite{Mizuta-NOY-2005pra}.
All of these methods 
 use the same basic concept -- 
 use the right combination of $E$ and $H$ fields so as to create a 
 pair of forward and backward-like fields.

Here I follow the most general formulation that I know of, 
 which is that of 
 Kinsler \cite{Kinsler-2010pra-dblnlGpm}, 
 as developed
 from earlier work \cite{Kinsler-RN-2005pra,Kinsler-2006arXiv-fleck}.
These handle the electric and magnetic properties of the 
 propagation medium on an equal footing, 
 incorporates the dispersive properties of the medium in a 
 very general way, 
 and retains all the vectorial behaviour of the fields.
The result is paired first-order equations for 
 the plane-polarized directional fields $G^\pm$ 
 (and a longitudinal component $G^\circ$).
Although complicated in the general case, 
 these simplify greatly in the usual case(s) of 
 transverse and/or paraxial propagation regimes. 
The cost of using these directional fields is that while we can 
 efficiently remove backward propagating contributions,
 computing the nonlinear terms is more demanding.
In contrast, 
 the work of Kolesik et al. and Mizuta et al. is distinguished 
 by a greater emphasis on the practical applications of 
 directional fields.

Because these new $G^\pm$ fields are directional, 
 we can efficiently separate out the forward-going part of the field, 
 and neglect the backward.
This is an important step, 
 because the standard Maxwell equations based approaches 
 treated in the previous section could not easily remove 
 the backward parts of the field, 
 and these can cause inconvenience in numerical simulations.
For example, 
 the spurious backward component caused by 
 imperfect initial conditions should no longer occur.


The definitions of the ${G}^{\pm}$ fields, 
 describing the transverse properties of a plane polarized EM field, 
 in the frequency domain are
~
\begin{eqnarray}
  \tilde{G}_x^{\pm} (\omega)
&=&
  \tilde{\alpha}_r (\omega)  \tilde{E}_x  (\omega)
 \pm 
  \tilde{\beta}_r (\omega)  \tilde{H}_y (\omega)
,
\label{eqn-S-defs-GvectorW}
\label{eqn-S-defs-Gvector}
\end{eqnarray}
The $\tilde{\alpha}_r$ and $\tilde{\beta}_r$ ``reference'' parameters 
 are best chosen to closely match the medium,
 whilst ignoring nonlinear effects, 
 so that
 $ \tilde{\alpha}_r(\omega) \tilde{\beta}_r(\omega) = 1 / c(\omega)$.
That is, 
 relevant (linear) dispersive properties 
 of the medium are included in the reference parameters, 
 i.e. that $\tilde{\alpha}_r(\omega) = \tilde{\epsilon}_r(\omega)^{1/2}$.
They have the definitions
~
\begin{eqnarray}
  \tilde{\epsilon} 
~~~~ = \tilde{\epsilon}_r(\omega) + \tilde{\epsilon}_c(\omega)
&=&
   \tilde{\alpha}_r^2(\omega) 
 + \tilde{\alpha}_r(\omega) ~ \tilde{\alpha}_c(\omega),
\label{eqn-defs-alphaX}
\\
  \tilde{\mu} 
~~~~ = \tilde{\mu}_r(\omega)  + \tilde{\mu}_c(\omega)
& =&
    \tilde{\beta}_r^2(\omega) 
  + \tilde{\beta}_r(\omega) ~ \tilde{\beta}_c(\omega),
\label{eqn-defs-betaX}
\end{eqnarray}
where the correction parameters $\tilde{\epsilon}_c$ and $\tilde{\mu}_c$
 represent the discrepancy between the true values and the reference.
These correction terms will 
 usually just be the nonlinearity.
More generally, 
 the smaller these correction terms are, 
 the better the match, 
 and the more likely it is that a description involving only ${G}^{+}$ 
 will suffice.
Note also that there are alternative ways of constructing 
 directional ${G}^{\pm}$-like fields \cite{Kinsler-2006arXiv-fleck}.

In the widely used moving frame defined by $v=1/\alpha_f \beta_f$, 
 where $\partial_{z} Q = \partial_{z'} Q - \alpha_f \beta_f \partial_t Q$, 
 using these $G_x^\pm$ fields 
 gives the (non-magnetic case) 
 propagation equation \cite{Kinsler-RN-2005pra}, 
~
\begin{eqnarray}
 -
  \partial_{z'} \tilde{G}_x^{\pm}
&=&
 \mp 
  \imath \omega 
  \tilde{\alpha}_r \tilde{\beta}_r 
  \left( 1 \mp \xi \right)
  ~
  \tilde{G}^{\pm}
\nonumber 
\\
&&
~~~~
 \mp 
  \frac{\imath \omega \tilde{\alpha}_c \tilde{\beta}_r}
       {2}
  \convol
    \left[ \tilde{G}_x^{+} + \tilde{G}_x^{-} \right]
 ~~
,
\label{eqn-firstorder-Gpm-comoving}
\end{eqnarray}
 where $\xi=\alpha_f \beta_f / \tilde{\alpha}_r \tilde{\beta}_r$.
Although this moving frame has no sensible limit as the frame speed
 tends to zero, 
 the stationary frame case can be recovered by setting $\xi=0$ 
 and replacing $z'$ by $z$.
$G^\pm$ field simulations usually assume $G_x^-=0$, 
 and treat only the forward traveling components  of the EM field.

Correctly writing down the form of nonlinear terms 
 for eqn. (\ref{eqn-firstorder-Gpm-comoving}) requires some care, 
 and consideration of the specific nonlinearity involved.
Fortunately the task is simplified because it is simply a rewriting of 
 the (electric) nonlinear term from Maxwell's equations with the 
 appropriate scaling factors relating $\alpha_c$ to $\epsilon$,
 and $G_x^\pm$ to $E$.

Wave equations with a more familiar appearance
 can be obtained using 
\begin{eqnarray}
  \tilde{E}^\pm(\omega) 
&=&
   \tilde{G}_x^\pm(\omega) / 2 \tilde{\alpha}_r(\omega)
.
\end{eqnarray}
These have the units of an electric field (i.e. V/m), 
 but actually incorporate information about the magnetic field as well.
If we take this step, 
 we can transform back into forward propagating 
 ``electric fields'' $E^\pm$, 
 and get
~
\begin{eqnarray}
 -
  \partial_{z'} \tilde{E}^{\pm}
&=&
 \mp 
  \imath \omega 
  \tilde{\alpha}_r \tilde{\beta}_r 
  \left( 1 \mp \xi \right)
  ~
  \tilde{E}^{\pm}
\nonumber 
\\
&&
~~~~
 \mp 
  \frac{\imath \omega \tilde{\alpha}_c \tilde{\beta}_r}
       {2}
  \convol
    \left[ \tilde{E}^{+} + \tilde{E}^{-} \right]
 ~~
.
\label{eqn-firstorder-Epm-comoving}
\end{eqnarray}

An approximate forward-only wave equation can be found
 by setting $E^-=0$ in eqn. (\ref{eqn-firstorder-Epm-comoving}),
 (or $G^-=0$ in eqn. (\ref{eqn-firstorder-Gpm-comoving})).
For a time response $\chi^{(3)}$ nonlinearity, 
 this is
~
\begin{eqnarray}
 -
  \partial_{z'}
  \tilde{E}^+(\omega)
&=&
 -
  \imath \omega 
  \tilde{\alpha}_r \tilde{\beta}_r 
  \left( 1 - \xi \right)
  ~
  \tilde{E}^+(\omega)
\nonumber
\\
&&
 -
  \imath \omega 
 \left\{
  \tilde{\beta}_r 
  \tilde{\epsilon}_c(\omega) 
  .
  \mathscr{F}
  \left[
    E_x^+(t)^2
  \right]
  (\omega) 
 \right\}
  \convol
   \tilde{E}^+(\omega)
\nonumber
\\
\label{eqn-Ep-chi3}
\end{eqnarray}

Notice the similarity to eqn. (\ref{eqn-pssd-dH}), 
 but that the field is propagated in a single first order equation, 
 rather than two 
 (i.e. both eqn. (\ref{eqn-pssd-dH}) and (\ref{eqn-pssd-dE})).
The cost is that it only propagates {\em forwards}, 
 but this is what we wanted.
Further, 
 the method can be implemented using 
 fewer Fourier transforms
 than are required for 
 a full Maxwell equation solver \cite{Kinsler-RN-2005pra}.
The gain is that of not solving for $\partial_z E$ (eqn. (\ref{eqn-pssd-dE})),
 which requires a pair of FFT's if done pseudospectrally. 
Solving for pulse propagating in a medium with dispersion 
 and a time dependent (or instantaneous) third order nonlinearity 
 therefore requires only six (or three) FFT's, 
 as compared to eight (or five) for solving Maxwell's equations.

However, 
 in practice the speed gain can be less clear cut.
A PSSD solver moves forward one full step $dz$ in two staggered steps, 
 one integrating for the magnetic field, 
 and integrating for  the electric field; 
 and only the magnetic field integration needs to calculate 
 the nonlinearity.
This staggered scheme is second order accurate even though 
 each stagger-step is only integrated using an Euler method.
We can achieve nearly the same level of accuracy for 
 the directional fields 
 by employing a leapfrog algorithm 
 \cite{Press-TVF-1992-numericalrecipies}.
If we wish to use more accurate (and so more complicated) 
 numerical integration algorithms
 (e.g. a Runga-Kutta scheme), 
 then we can only outperforms the staggered (or leapfrog)
 PSSD  schemes if the propagation step size is (greater than) twice 
 that of the staggered PSSD.
To complicate matters further, 
 for reasons of numerical stability, 
 we often need to 
 tie the propagation step $dz$ to the time grid step $dt$.
This means that if there are bandwidth constraints limiting our $dt$, 
 we may not have as much much freedom to adjust $dz$ as we might like.

%
\subsubsection{Special case: $\chi^{(2)}$}

In the case of a $\chi^{(2)}$ nonlinearity, 
 two different field polarizations are coupled together, 
 and the equations given above tend to obscure the final form the 
 nonlinear term will take.
In this case, 
 the time-domain displacement fields $D$ in the two polarizations are
~
\begin{eqnarray}
  D_x
&=&
  \epsilon_x \convol E_x 
 +
  2 \epsilon_0 \chi^{(2)} E_x E_y
~~~~
=
 \epsilon_x \convol E_x + \mathscr{N}^{(2)}_x
,
\\
  D_y
&=&
  \epsilon_y \convol E_y
 +
  \epsilon_0 \chi^{(2)} E_x^2
~~~~
=
 \epsilon_x \convol E_x + \mathscr{N}^{(2)}_y
.
\end{eqnarray}
If we assume that all of the linear response of the material 
 (denoted above by $\epsilon_x, \epsilon_y$) is 
 absorbed into the reference parameters $\tilde{\alpha}_r, \tilde{\beta}_r$, 
 we need only consider the nonlinear part.
Note in particular that for
 the $D_y$ field (i.e. $\mathscr{N}^{(2)}_y$)
 this does not depend on $E_y$, 
 meaning that the forms of the wave equations given above 
 (aimed largely at a $\chi^{(3)}$ system)
 are not very useful.

First, 
 note that for a $\chi^{(3)}$ nonlinearity
~
\begin{eqnarray}
  \mathscr{N}^{(3)}
&=&
  \epsilon_0 \chi^{(3)} E^3
,
\\
  \tilde{\mathscr{N}}^{(3)}
&=&
  \mathscr{F}\left[\epsilon_0 \chi^{(3)} E^2\right]
  \convol E
\\
&=&
  \tilde{\alpha}_r \tilde{\alpha}_c 
  \convol
    \tilde{E}
,
\end{eqnarray}
and these give a nonlinear term for the wave equations of
~
\begin{eqnarray}
  \frac{\imath \omega \tilde{\alpha}_c \tilde{\beta}_r}
       {2}
  \convol
    \left[ \tilde{G}_x^{+} + \tilde{G}_x^{-} \right]
&=&
  \imath \omega \tilde{\beta}_r . 
  \tilde{\alpha}_r \tilde{\alpha}_c 
  \convol
    \tilde{E}
,
\end{eqnarray}

By comparing these $\chi^{(3)}$ terms, 
 we can see that in the $\chi^{(2)}$ case, 
 the nonlinear terms in the $\tilde{G}_x^{\pm}$ 
 and $\tilde{G}_y^{\pm}$ wave equations 
 (see eqn. (\ref{eqn-firstorder-Gpm-comoving}))
 will be rewritten as follows
~
\begin{eqnarray}
x: ~~~~ 
  \frac{\imath \omega \tilde{\alpha}_c \tilde{\beta}_r}
       {2}
  \convol
    \left[ \tilde{G}_x^{+} + \tilde{G}_x^{-} \right]
&\Rightarrow&
  \imath \omega \tilde{\beta}_r 
  \mathscr{F}\left[2 \epsilon_0 \chi^{(2)} E_x E_y\right]
, ~~~~
\label{eqn-gpm-chi2-Gx}
\\
y: ~~~~ 
  \frac{\imath \omega \tilde{\alpha}_c \tilde{\beta}_r}
       {2}
  \convol
    \left[ \tilde{G}_y^{+} + \tilde{G}_y^{-} \right]
&\Rightarrow&
  \imath \omega \tilde{\beta}_r 
  \mathscr{F}\left[\epsilon_0 \chi^{(2)} E_x^2\right]
. ~~~~
\label{eqn-gpm-chi2-Gy}
\end{eqnarray}
These can then be put in a form containing only
 $\tilde{G}_x^{\pm}, \tilde{G}_y^{\pm}$ if desired, 
 but it is simplest to reconstruct the $E_x, E_y$
 directly before calculating the nonlinear terms.
If a more extensive collection of the $\chi^{(2)}$ coefficients 
 needs to be included, 
 this procedure can be reproduced
 using the appropriate nonlinear field combinations.
Further, 
 if the time-response of the nonlinearity is also important, 
 then we can include this by replacing
 $\chi^{(2)} E_x E_y$ and $\chi^{(2)} E_x^2$ with appropriate convolutions:
 e.g. $(\chi^{(2)} \convol E_y ) E_x$ and $(\chi^{(2)} \convol E_x) E_x$.

For the $\tilde{E}^{\pm}$-like
 wave eqns. (\ref{eqn-firstorder-Epm-comoving})
 the nonlinear terms are
~
\begin{eqnarray}
x: ~~~~ 
  \frac{\imath \omega \tilde{\alpha}_c \tilde{\beta}_r}
       {2}
  \convol
    \left[ \tilde{E}_x^{+} + \tilde{E}_x^{-} \right]
&\Rightarrow&
  \frac{\imath \omega \tilde{\beta}_r}{2 \tilde{\alpha}_r}
  \mathscr{F}\left[2 \epsilon_0 \chi^{(2)} E_x E_y\right]
, ~~~~
\label{eqn-gpm-chi2-Ex}
\\
y: ~~~~ 
  \frac{\imath \omega \tilde{\alpha}_c \tilde{\beta}_r}
       {2}
  \convol
    \left[ \tilde{E}_y^{+} + \tilde{E}_y^{-} \right]
&\Rightarrow&
  \frac{\imath \omega \tilde{\beta}_r}{2 \tilde{\alpha}_r}
  \mathscr{F}\left[\epsilon_0 \chi^{(2)} E_x^2\right]
. ~~~~
\label{eqn-gpm-chi2-Ey}
\end{eqnarray}

Since we will want to apply the nonlinear effects in the time domain, 
 we need to back-transform the terms in 
 eqns. (\ref{eqn-gpm-chi2-Gx},\ref{eqn-gpm-chi2-Gy})
 or eqns. (\ref{eqn-gpm-chi2-Ex},\ref{eqn-gpm-chi2-Ey}),
 requiring a pair of Fourier transforms
 in addition to those required to get the time-domain fields.
If using the $\tilde{G}^{\pm}$ form, 
 there is an additional transform, 
 because we also need the time domain field(s) $E(t)$ --
 with the $\tilde{E}^{\pm}$ form $E(t)$ can be found directly.

Further simplifications can be made: 
 e.g. in a semi-wideband limit around a central frequency $\omega_0$,
 we can assume the frequency dependence
 of the $\tilde{\alpha}$ parameters in the nonlinear terms vanishes, 
 so that the transform(s) to convert from $G^\pm$ to $E$ is unecessary.
In an SVEA-like narrowband limit all these transforms 
 vanish because (in the nonlinear terms) the frequency dependence
 of the $\tilde{\alpha}$ parameters vanish and the the 
 factor of $\omega$ simple becomes $\omega_0$.

%
\subsection{Envelopes}\label{Ss-directional-envelopes}

Here I have intentionally simplified the definitions to best 
 match what is most likely to be used in practice:
 a forward propagating $G^+$ (or $E^+$) only model.
A more complete description of $G^\pm$ envelopes, 
 such as that in \cite{Kinsler-RN-2005pra},
 would include the role of forward and backward traveling 
 envelopes for both of $G^\pm$.

We have seen that in the forward-only approximation, 
 $G^+$ and $E^+$ follow identical equations of motion.
The envelope and carrier representation of $E^+$ is
~
\begin{eqnarray}
  E^+(t;z)
&=&
  C(t;z)
  e^{\imath\left(k_0 z-\omega_0 t\right)} 
 +
  C^*(t;z)
  e^{-\imath\left(k_0 z-\omega_0 t\right)} 
.
\label{eqn-Gp-env}
\end{eqnarray}

I do not apply this envelope definition to the general equation of motion
 because how $\tilde{\alpha}_c$ is expressed depends on the field and
 therefore on those envelopes.

We now, for the case of a time response $\chi^{(3)}$ nonlinearity,
 substitute eqn. (\ref{eqn-Gp-env})
 into eqn. (\ref{eqn-Ep-chi3}),
 we then (as usual) 
 split the normal and c.c. parts,
 cancel exponentials,
 and rearrange leaving only the $\partial_z$ terms on the left,
~
\begin{eqnarray}
 -
  \partial_{z'} \tilde{C}(\omega)
&=&
 - 
  \imath \omega
  \left( 1 - \xi \right)
    \tilde{\beta}_r \tilde{\alpha}_r \tilde{C}(\omega)
 +
  \imath k_0 \tilde{C}(\omega)
\nonumber
\\
&&
 -
  \imath \omega 
  \tilde{\beta}_r 
  \tilde{\epsilon}_c(\omega+\omega_0) 
  .
  \mathscr{F}
  \left[
    2 \left| C(t) \right|^2
  \right]
  (\omega)
  .
  \tilde{C}(\omega)
\nonumber
\\
&&
 ~
 -
  \imath \omega 
  \tilde{\beta}_r 
  \tilde{\epsilon}_c(\omega+\omega_0) 
  .
  \mathscr{F}
  \left[
    C(t)^2
  \right]
  (\omega)
  .
  \tilde{C}^*(\omega)
\nonumber
\\
&&
 ~~
 -
  \imath \omega 
  \beta_r 
  \tilde{\epsilon}_c(\omega+3\omega_0) 
  .
  \mathscr{F}
  \left[
     C(t)^2
  \right]
  (\omega)
  .
  \tilde{C}(\omega)
,
\nonumber
\\
\label{eqn-Ap-chi3}
\end{eqnarray}
The first line on the RHS will mostly cancel in the narrowband case, 
 since $\tilde{\beta}_r \tilde{\alpha}_r = 1/c(\omega)$, 
 and $k_0 = \omega_0 / c(\omega_0)$, 
 thus with $\delta = \omega-\omega_0$ it becomes
~
\begin{eqnarray}
 - 
  \imath 
  \left[
    \frac{\omega}{c(\omega)}
   -
    k_0
  \right]
&=&
 - 
  \imath 
  \left[
    k(\omega) - k_0 
  \right]
\nonumber
\\
&=&
 - 
  \imath 
  \left[
    \left.
      \frac{\partial k}
           {\partial \omega}
    \right| _{\omega_0}
    \delta
  +
      \frac{1}{2}
    \left.
      \frac{\partial^2 k}
           {\partial \omega^2}
    \right| _{\omega_0}
    \delta^2
  +
    ...
  \right]
,
 ~~~~
\end{eqnarray}
 where in the truncated expansion on the second line 
 we can see the expected group velocity and 
 group velocity dispersion terms.

Note that eqn. (\ref{eqn-Ap-chi3}) is directly comparable
 to one derived from the NEE of Brabec and Krausz \cite{Brabec-K-1997prl}, 
 but the {\em only} approximation I have made is to discard
 backward propagating fields.
Since the NEE makes several additional approximations, 
 eqn. (\ref{eqn-Ap-chi3}) is {\em more accurate} and 
 {\em less approximate}.
Indeed, 
 Brabec and Krausz were fortunate in that their chosen 
 approximations produced a result remarkably similar 
 to that from the less restricted directional fields approach.
Note that Kolesik and Moloney \cite{Kolesik-M-2004pre} also reduced 
 their directional wave equation to a number of special cases, 
 including that of Brabec and Krausz.

%
\subsection{Transverse effects}\label{Ss-directional-transverse}

As for Maxwell's equations, 
 there are two main transverse effect likely to be of interest 
 in pulse propagation models: 
 mode averaging, 
 and diffraction or off-axis propagation.

Mode averaging is easy to incorporate if 
 you assume some known transverse profile for the mode: 
 e.g. for an optical fibre or some other waveguide.
The transverse derivatives vanish, 
 and the material properties are evaluated as an 
 integral over the transverse dimensions, 
 weighted by the mode function.
This is just the same as for Maxwell's equations, 
 although we now may be averaging slightly different 
 quantities 
 (e.g. $\alpha_r$ rather than $\epsilon$).
The work of
 Kolesik et al. \cite{Kolesik-MM-2002prl,Kolesik-M-2004pre}
 allows for transverse mode structure, 
 that of Mizuta et al. \cite{Mizuta-NOY-2005pra}
 for transverse averaging over a single mode.

Diffraction and off-axis propagation are again much harder to 
 understand, 
 because (again)
 they result from a coupling between the vector components of 
 the $E$ and $H$ fields -- 
 including those along the propagation direction.
However, 
 second order wave equations derived from 
 the first order directional fields equations 
 exhibit a $\nabla_\perp^2$ diffraction term
 which is the same as that seen
 in standard second order wave equations
 (see e.g. eqn. (\ref{exact-BKP}), 
 in section \ref{S-secondorder}).
This means that weakly transverse effects can be accurately 
 incorporated by 
 using a split step scheme alternating between the wave equation
 and a $\nabla_\perp^2$ diffraction term.
Note that 
 Kolesik et al. \cite{Kolesik-M-2004pre} 
 had  wave equations incorporating 
 diffraction (transverse) terms.

%
\section{Second order wave equations}\label{S-secondorder}

The standard second order wave equation applies to propagation 
 in non-magnetic materials.
If we consider the case of small transverse
 inhomogeneities of the polarization, 
 the three dimensional wave equation in typical notation 
 (e.g. from \cite{Brabec-K-1997prl,Kinsler-N-2003pra}) is 
~
\begin{eqnarray}
  \left( \partial_z^2 + \nabla_\bot^2 \right) E(\vec{r},t)
- \frac{1}{c^2}
  \partial_t^2
  \left\{
    \epsilon_L(\tau) \convol E(\vec{r},t)
  \right\}
=
  \frac{4\pi}{c^2}
  \partial_t^2
  P_{nl}(\vec{r},t)
.
\nonumber
\\
\label{eqn-3DWE-time}
\end{eqnarray}


Here $\nabla_\bot^2$ is the transverse Laplace operator,
 $\epsilon_L(t) = (2\pi)^{-1} \int_{-\infty}^\infty d\omega
   \tilde{\epsilon}_L(\omega) e^{\imath \omega t}$, 
 $\tilde{\epsilon}_L(\omega) = 1 + 4\pi \chi(\omega)$, 
 and $\chi(\omega)$ is the linear electric susceptibility. 
The electric field $E$ propagates along the $z$ direction. 
Both $E$ and the nonlinear polarization $P_{nl}$
 are polarized parallel to the $x$ axis.

Because of their starting point, 
 methods based on this second order equation are
 slightly more restricted than those starting from Maxwell's equations.
However, 
 the differences in practice will likely be small, 
 especially in the usual case of non-magnetic propagation media.

Most uses of eqn. (\ref{eqn-3DWE-time}), 
 notably the slowly varying envelope approximation (SVEA) 
 relies on using an envelope-carrier description for the fields, 
 then expands for weak dispersion, 
 and resonant nonlinear perturbations about this carrier.
This approach is discussed below in subsection \ref{Ss-secondorder-trad}.

Alternatively, 
 we can attempt to factorise the equation into a product of 
 two first order parts, 
 as can be done for linear waves (see e.g. \cite{Tanuiti-Nishihara-NLW}).
Factorization 
 is considerably more useful than the traditional approach, 
 and is discussed below in subsection \ref{Ss-secondorder-factor}.

%
\subsection{Traditional approach}\label{Ss-secondorder-trad}

Unlike the other approaches discussed in this paper, 
 the traditional approach assumes the use of an 
 envelope-carrier description of the field.

Kinsler and New \cite{Kinsler-N-2003pra,Kinsler-2002-fcpp}
 presented a comprehensive re-derivation 
 of the envelope propagation equation based on the 
 second order wave equation, 
 which subsumes the SVEA
 and Brabec and Krausz's NEE \cite{Brabec-K-1997prl}
 as special cases.
Since it is the most general, 
 I use the Kinsler and New calculation,
 leaving some definitions to their paper rather than repeat them here.
Noting that $\xi$ and $\tau$ are scaled space and time variables, 
 that $alpha$ and $\beta$ have different meanings from the rest of this paper, 
 and that $\hat{D}'$ contains the dispersion terms, 
 we have
~
\begin{eqnarray}
&&\partial_\xi
  A(\vec{r}_\bot,\xi,\tau) 
\nonumber
\\
&=&
    \left( 
    - \frac{ \alpha_0}{\beta_0}
    + \imath  \hat{D}'
    \right)
  A(\vec{r}_\bot,\xi,\tau) 
  + 
    \frac{\left( \imath / 2 \beta_0^2 \right) \nabla_\bot^2}
         { \left( 1 + \imath \sigma \partial_\tau \right)}
    A(\vec{r}_\bot,\xi,\tau) 
\nonumber
\\
&+&
    \frac{2 \imath \pi }{n_0^2}
    \frac{\left(1 + \imath \partial_\tau \right)^2}
         {\left( 1 + \imath \sigma \partial_\tau \right)}
    B(\vec{r}_\bot,\xi,\tau ; A)
+
    \frac{ T_{R} }
         { 1 + \imath \sigma \partial_\tau }
,
\label{exact-BKP}
\end{eqnarray}
where 
~
\begin{eqnarray}
T_{R} 
&=& 
  \left[
  -
    \frac{\imath q^2 }{2}
    \partial_\xi^2
  +
    \frac{\imath}{2}
    \left(
      \frac{ \alpha_0}{\beta_0}
    - \imath  \hat{D}'
    \right)^2
  \right]
    A(\vec{r}_\bot,\xi,\tau) 
.
\label{exact-BKP-Tr}
\end{eqnarray}

Eqn. (\ref{exact-BKP}) {\em is exact} -- it contains no more approximations
than the starting point eqn. (\ref{eqn-3DWE-time}) except for the expansion 
of $\epsilon$ in powers of $\omega$.  
If we set  $T_{R}=0$, this gives us a 
 generalized few cycle envelope (GFEA) equation, 
 which contains the SVEA \cite{Shen-PNLO}.
Brabec and Krausz's NEE can be recovered from 
 eqn. (\ref{exact-BKP}) in the 1D case 
 where phase and group velocities are the same (i.e. $\sigma=1$), 
 likewise Porras's SEEA \cite{Porras-1999pra} can be identified in 
 the diffraction term.
Of course we cannot just set the $T_{R}$  term to zero without
 some justification, but this has already been extensively 
 discussed,
 not only in both \cite{Kinsler-N-2003pra}, 
 but also the detailed analysis \cite{Kinsler-2002-fcpp}.

Now consider the complicated few-cycle correction to 
 the polarization term in eqn. (\ref{exact-BKP}), 
 which
 contains partial derivatives ($1+\imath \sigma \partial_\tau$) in the
 denominators.
These will need to be evaluated by Fourier transforming into the
 conjugate frequency space ($\Omega$).
Further, 
 the $T_{R}$ term is divided by another such term.
Clearly these might, 
 in wideband cases, 
 result in denominators close to zero, 
 causing the approximations to fail.
This means they put a serious brake on the validity of 
 {\em any} such approach, 
 especially if the bandwidth of the pulse approaches the carrier frequency.

Note that the best first order expansion of the few-cycle corrections 
 to the polarization term is more general than that given by
 Brabec and Krausz, 
 and contains the group to phase velocity ratio $\sigma$, 
 i.e.
~
\begin{eqnarray}
    \frac{2 \imath \pi }{n_0^2}
    \frac{\left(1 + \imath \partial_\tau \right)^2}
         {\left( 1 + \imath \sigma \partial_\tau \right)}
    B(\xi,\tau ; A)
&\approx&
    \frac{2 \imath \pi }{n_0^2}
    \left(1 + \imath \sigma  \partial_\tau \right)
    B(\xi,\tau ; A)
, 
~~~~
\label{eqn-betterthanBK}
\end{eqnarray}

Unfortunately for the venerable SVEA based on the 
 second order wave equation, 
 and even its most general variant presented here, 
 the directional fields method discussed in the previous section 
 \ref{S-directional}
 has made it utterly redundant; 
 as, 
 indeed, 
 has the approach in 
 the following subsection \ref{Ss-secondorder-factor}.
There is no reason to use any form of the GFEA or SVEA 
 when we can generate equations like 
 eqns. (\ref{eqn-firstorder-Epm-comoving},\ref{eqn-Ep-chi3})
 by not only using {\em fewer} approximations, 
 but much {\em simpler} ones than those taken by 
 neglecting $T_R$.

%
\subsection{Time propagated direct solution}\label{Ss-secondorder-time}

It is of course possible to solve the second order wave equation 
 by propagating it in time, 
 either with or without the use of an envelope and carrier.
This approach has been used with significant success by 
 Scalora and co-workers 
 (e.g. their early work 
 \cite{Dowling-SBB-1994jap,Scalora-DBB-1994prl,Scalora-C-1994oc}).
By propagating in time reflections are handled correctly, 
 an important feature when treating structured materials.
Generally the solution is achieved retaining the second order spatial
 derivatives (both in $z$ and transversely in $x,y$), 
 but approximating the time derivatives to first-order.
The approximation is made using 
 an envelope with a well-chosen carrier frequency, 
 and gives rise to the SVEAT, 
 or slowly varying envelope approximation in {\em time}.

%
\subsection{Short pulse equation (SPE)}\label{Ss-secondorder-spe}

The second order wave equation 
 can be converted into the SPE by using a multiscale expansion
 \cite{Schafer-W-2004pd}.
First, 
 specialize to a third-order nonlinearity (strength $p$)
 and only second order (ordinary) dispersion (strength $d$)
 and then rewrite the second order wave equation as
~
\begin{eqnarray}
  \partial_z^2
    E(t;z)
 -
  \frac{1}{c_1^2}
  \partial_t^2
    E(t;z)
 -
  d_2
    E(t;z)
 -
  p
  \partial_t^2
    E(t;z)^3
&=&
   0. ~~~~
\label{eqn-2ndWave-spe-start}
\end{eqnarray}

We introduce the scaled co-moving frame variables
 $\tau = (t-z/c)/\sigma$ 
  so that $\partial_t=(1/\sigma)\partial_\tau$, 
 and $z_n = \sigma^n z$ 
  so that $\partial_z = -(1/c_1\sigma)\sigma^ n\partial_{z_n}$; 
hence after simplification eqn. (\ref{eqn-2ndWave-spe-start}) becomes
~
\begin{eqnarray}
 -
  \frac{2}{c_1}
  \partial_\tau \partial_{z_1}
    E(t;z)
 -
  d_2
    E(t;z)
 -
  \frac{p}{\sigma^2}
  \partial_\tau^2
    E(t;z)^3
&=&
   0. ~~~~
\label{eqn-2ndWave-spe-scaled}
\end{eqnarray}

Now, 
 writing the field in multiscaled form as a power series in 
 components $E_i$ scaled by factors of $\sigma$, 
 we have
~
\begin{eqnarray}
  E(t;z)
&=&
  \sigma 
  E_0(\tau, z_1, z_2, ...)
 +
  \sigma^2
  E_1(\tau, z_1, z_2, ...)
 +
  ...
\end{eqnarray}
and to leading order, 
 we can write eqn. (\ref{eqn-2ndWave-spe-scaled}) down 
 as the SPE
~
\begin{eqnarray}
 -
  \frac{2}{c_1}
  \partial_\tau \partial_{z_1}
    E_0
 -
  d_2
    E_0
 -
  p
  \partial_\tau^2
    E_0^2
&=&
   0. ~~~~
\label{eqn-2ndWave-spe}
\end{eqnarray}

This equation has spawned a literature all of its own, 
 because it (like the ordinary nonlinear Schr\"odinger equation)
 provides a rich variety of mathematical solutions.
Note that what is essentially a variant of the SPE, 
 but specialized for HHG by generalizing the dispersion and nonlinearity
 is also in use \cite{Geissler-TSSKB-1999prl}.

%
\subsection{Factorization approach}\label{Ss-secondorder-factor}

An alternative to the traditional style of derivation discussed above, 
 we can instead factorise the second order wave equation in a 
 way similar to that done for linear waves 
 (see e.g. \cite{Tanuiti-Nishihara-NLW}).
This was initially suggested by Shen \cite{Shen-PNLO}, 
 followed by Blow and Wood \cite{Blow-W-1989jqe},  
 and more recently revisited by Ferrando et al. \cite{Ferrando-ZCBM-2005pre}
 and Genty et al. \cite{Genty-KKD-2007oe}; 
 the most general formulation, 
 which also allows for magnetic effects is at \cite{Kinsler-2010pra-fchhg}.

Note that the work of Weston examines this kind of wave-splitting
 with more mathematical rigour
 (see e.g. \cite{Weston-1993jmp}), 
 although without consideration of residual terms, 
 and (at least initially) in the context of reflections and scattering.
This theory was based on that from the earlier work of 
 Beezley and Krueger \cite{Beezley-K-1985jmp} 
 who applied wave-splitting concepts to optics.

First we reduce eqn. (\ref{eqn-3DWE-time}) to the 1D strictly paraxial limit; 
 then transform into frequency space.
Here I re-use the symbol $\beta$ as the propagation wave vector 
 to match the notation of Genty et al. \cite{Genty-KKD-2007oe}, 
 so that $\beta(\omega) = \omega \sqrt{\epsilon_r(\omega) \mu_0}$.
The wave equation therefore is
~
\begin{eqnarray}
  \partial_z^2
    E(t;z)
 -
  \frac{1}{c^2}
  \partial_t^2
    E(t;z)
 -
  \mu_0
  \partial_t^2
    P(t;z)
&=&
   0, ~~~~
\\
  \nabla^2
    \tilde{E}(\omega;z)
 +
  \beta(\omega)^2
    \tilde{E}(\omega;z)
 +
  \mu_0
  \omega^2
    \tilde{P}(\omega;z)
&=&
   0, ~~~~
\label{eqn-2ndWave-simple}
\\
  \partial_z^2
    \tilde{E}(\omega;z)
 +
  \beta^2(\omega)
    \tilde{E}(\omega;z)
 +
  \beta^2(\omega)
  \tilde{\mathscr{N}}
   \convol 
    \tilde{E}(\omega;z)
&=&
   0, ~~~~
\label{eqn-2ndWave}
\end{eqnarray}
where for a third order nonlinearity, 
 with  $\epsilon_c = \epsilon_0 \chi^{(3)}$,
~
\begin{eqnarray}
  \mathscr{N}
&=&
  \mu_0 \epsilon_0 \chi^{(3)} \omega^2 E(r,t;z)^2 / \beta(\omega)^2
\\
&=&
  \mu_0 \epsilon_c \omega^2 E(r,t;z)^2 / \beta(\omega)^2
\\
&=&
  \frac{\chi^{(3)} }{n(\omega)^2} E(r,t;z)^2
.
\end{eqnarray}

I now briefly consider three factorization approaches, 
 from the simple method of Blow and Wood \cite{Blow-W-1989jqe}, 
 an improved version, 
 and finally the most rigorous approach.
Although these traditionally involve an envelope-carrier decomposition
 introduced early in that calculation
 (see  Blow and Wood), 
 the step is in fact unnecessary and I omit it.

\subsubsection{Simple factorization}

Factorization approaches are simple in two situations: 
 a dispersionless medium with an instantaneous nonlinearity, 
 and a dispersive medium with no nonlinearity.
In the dispersionless nonlinearity case, 
 we can factorise in the time domain.
In the linear dispersive case, 
 we can factorise in the frequency domain.
In the dispersive nonlinear case, 
 it is (usually) not possible to analytically factorise 
 the second order wave equation.


\noindent
\textbf{Basic Blow and Wood: }
The simplest, 
 but least rigorous method of factorising is that 
 of Blow and Wood \cite{Blow-W-1989jqe}.
Ignoring many of mathematical difficulties, 
 Blow and Wood ignored the details of nonlinearity and dispersion.
Remembering that $\beta = \beta(\omega)$,
 and without their envelope-carrier decomposition, 
 they had 
~
\begin{eqnarray}
  \left[
    \partial_z
   +
    \imath
      \beta
    \sqrt{
      1
     +
      \tilde{\mathscr{N}} \convol
    }
  \right]
  \left[
    \partial_z
   -
    \imath
      \beta
    \sqrt{
      1
     +
      \tilde{\mathscr{N}} \convol
    }
  \right]
  \tilde{E}
&=&
  0
.
\end{eqnarray}
They then separated out the forward propagating term.
The envelope equivalent of this was then 
 expanded using a ``weak nonlinearity'' assumption
 with a binomial expansion, 
 keeping only the first order corrections.


\noindent
\textbf{Improved Blow and Wood: }
The approach of Blow and Wood ignores the mathematical difficulties
 due to the use of the square root
 in combination with the frequency-domain convolutions
 between the nonlinear term $\mathscr{N}$ and 
 the field spectrum $\tilde{E}$;

Fortunately, 
 we can instead ``complete the square''
 (e.g. $1+N \simeq 1+N+N^2/4 = \left(1+N/2\right)^2$), 
 enabling us to preserve the convolutions correctly.
This requires us to make a weak nonlinearity approximation,
 but it is one nearly identical to that used when expanding
 the square root in the Blow and Wood calculation.
So, 
 with a the weak nonlinearity constraint 
~
\begin{eqnarray} 
  \frac{1}{2}
  \tilde{\mathscr{N}} \convol \tilde{E}
\ll
  1
,
\end{eqnarray}
we get 
~
\begin{eqnarray}
  \left[
    \partial_z
   +
    \imath
      \beta
    \left(
      1
      +
    \frac{\tilde{\mathscr{N}} \convol}{2}
    \right)
  \right]
  \left[
    \partial_z
   -
    \imath
      \beta
    \left(
      1
      +
    \frac{\tilde{\mathscr{N}} \convol}{2}
    \right)
  \right]
    \tilde{E}
&=&
  0
.
\label{eqn-factored-2nd-nlLeft}
\end{eqnarray}

By assuming the forward-like and backward-like terms in square 
 brackets factorise, 
~
\begin{eqnarray}
  \left[
    \partial_z
   \pm
    \imath
      \beta
    \left(
      1
      +
    \frac{\tilde{\mathscr{N}} \convol}{2}
    \right)
  \right]
    \tilde{E}
&=&
  0
.
\label{eqn-factored-2nd-nlLeft-fwd}
\\
    \partial_z
    \tilde{E}
&=&
 \pm
    \imath
      \beta
      \tilde{E}
 ~~
 \pm
    \imath
      \beta
    \frac{\tilde{\mathscr{N}} \convol}{2}
      \tilde{E}
.
\end{eqnarray}

While this equation can give excellent results, 
 it is restricted to weak nonlinearity:
 as we see below, 
 it lacks the nonlinear coupling term
 between the forward and backward propagating fields.

\subsubsection{Linear factorization}

Kinsler \cite{Kinsler-2010pra-fchhg} treats this approach in detail, 
 separating this second order equation into two first order equations, 
 using a method based on Ferrando et al.'s \cite{Ferrando-ZCBM-2005pre} 
 application of Greens functions.
This follws early applications of factorization 
 to nonlinear waveguides, 
 such as that by Genty et al. \cite{Genty-KKD-2007oe}
 with their nonlinear envelope equation.

The first step to achieving a first order wave equation containing
 the necessary physics but without unnecessarily complex approximations
 is to rewrite the wave eqn. (\ref{eqn-3DWE-time})
 to emphasize those contributions that, 
 without any coupling, 
 would freely propagate forward and backwards respectively.
To do this choose
 a specific propagation direction (e.g. along the $z$-axis), 
 and then denote the orthogonal components (i.e. along $x$ and $y$) 
 as transverse behaviour.
The wave equation eqn. (\ref{eqn-2ndWave}) can then be written
~
\begin{eqnarray}
  \left[
    \partial_z^2 
   +
    \frac{n^2(\omega) \omega^2}{c^2}
  \right]
  E(\omega)
&=&
 -
  \mathscr{Q}
.
\label{eqn-factored-2nd-nlRight-start}
\end{eqnarray}
Here I have moved some or all of the linear response 
 (e.g. the refractive index)
 out of the total polarization, 
 and over to the LHS as $n^2(\omega)$.
The remaining polarization term $\mathscr{Q}$ 
 would then include any nonlinearity
 (e.g. $\tilde{\mathscr{N}} \convol \tilde{E}$)
 or atomic response; 
 the diffraction
 (i.e. $\nabla_\perp^2 E$); 
 and indeed (if desired) even some linear terms 
 such as the angular dependence of the refractive index.
After Fourier transforming $z$ into $k$-space, 
 where the $\partial_z$ becomes $-\imath{k}$,
 we have
~
\begin{eqnarray}
  \left[
   -  k^2
   +
      \beta^2
  \right]
    \tilde{E}
&=&
 -
  \mathscr{Q}
\\
    \tilde{E}
&=&
  \frac{1}
       {k^2 - \beta^2}  
    \mathscr{Q}
\quad
=
    \frac{1}
         {
          \left(k+\beta\right)
          \left(k-\beta\right)
          }
    \mathscr{Q}
\\
  \tilde{E}_+ + \tilde{E}_-
&=&
 -
  \frac{1}{2\beta}
  \left[
    \frac{1}
         {k+\beta}
   -
    \frac{1}
         {k-\beta}
  \right]
    \mathscr{Q}
.
~~~~
\label{eqn-factored-2nd-nlRight-total}
\end{eqnarray}
where $E$ is now written as a sum of both forward and backward 
 propagating parts $\tilde{E} = \tilde{E}_+ + \tilde{E}_-$.
I now split eqn. (\ref{eqn-factored-2nd-nlRight-total})
 into a sum of two parts, 
 where each half represents the propagation
 of the forward field $E_+$ or the backward field $E_-$,
 and rearrange, 
~
\begin{eqnarray}
  \tilde{E}_\pm
&=&
  \pm
    \frac{1 / 2\beta}
         {k \mp \beta}
    \mathscr{Q}
\\
  \left[
    k \mp \beta
  \right]
  \tilde{E}_\pm
&=&
 \pm
  \frac{1}{2\beta}
    \mathscr{Q}
.
\end{eqnarray}
Now I transform back from $k$-space into $z$, 
 and multiply by $\imath$, 
 so that 
~
\begin{eqnarray}
  \left[
    \partial_z \mp \imath \beta
  \right]
  \tilde{E}_\pm
&=&
 \pm
  \frac{\imath}{2\beta}
    \mathscr{Q}
\\
    \partial_z
  \tilde{E}_\pm
&=&
  \pm
  \imath \beta
  \tilde{E}_\pm
 \pm
  \frac{\imath}{2\beta}
    \mathscr{Q}
.
\label{eqn-factored-2nd-nlRight-fwdbck}
\end{eqnarray}

If our polarization $\mathscr{Q}$ contains 
 a nonlinearity $\beta^2 \tilde{\mathscr{N}} \convol \tilde{E}$
 and diffraction terms $\nabla_\perp^2 E$
 we have \cite{Kinsler-2010pra-fchhg}
\begin{eqnarray}
    \partial_z
  \tilde{E}_\pm
&=&
  \pm
  \imath \beta
  \tilde{E}_\pm
  \pm
   \frac{\imath \beta}{2}
  \tilde{\mathscr{N}}
 \convol
  \left[
    \tilde{E}_+ + \tilde{E}_-
  \right]
 \pm
  \frac{\imath}{2\beta}
    \nabla_\perp^2 
  \left[
    \tilde{E}_+ + \tilde{E}_-
  \right]
. \quad
\label{eqn-factored-2nd-nlRight-fwdbck2}
\end{eqnarray}

If we compare this result
 (i.e. eqn. (\ref{eqn-factored-2nd-nlRight-fwdbck}))
 with the comparable equations for 
 the directional fields $G^\pm$,
 in particular with the electric field form given in 
 eqn. (\ref{eqn-firstorder-Epm-comoving}); 
 we see that they are essentially identical:
 since $\omega \alpha_r \beta_r = \omega / c_r \leftrightarrow \beta$.

A similar procedure can be applied to eqn. (\ref{eqn-factored-2nd-nlLeft})
 if desired.
Eqn. (\ref{eqn-factored-2nd-nlRight-fwdbck}) is almost the same as 
 the (more approximate) eqn. (\ref{eqn-factored-2nd-nlLeft-fwd}).
Since the RHS nonlinear term is a function of $(E_++E_-)$, 
 it provides a route for coupling between the forward and 
 backward waves; 
 its form can be obtained from the nonlinear part of (e.g.) 
 eqn (\ref{eqn-pssd-dH}).
A specific example for the case of a time-response $\chi^{(3)}$
 nonlinearity has been given in \cite{Genty-KKD-2007oe}, 
 but in my notation it is identical to that for the 
 rescaled directional $G^\pm$ fields (i.e. $E^\pm$), 
 i.e. eqn. (\ref{eqn-Ep-chi3}).

%
\subsubsection{Special case: $\chi^{(2)}$}

In the case of a $\chi^{(2)}$ nonlinearity, 
 two different field polarizations are coupled together, 
 and the equations given above tend to obscure the final form the 
 nonlinear term will take.
First, 
 note that the factorisation process changes the nonlinear term from
 $\beta^2 \omega^2 \tilde{\mathscr{N}} \convol \tilde{E}$
 into $\imath \beta \omega^2 \tilde{\mathscr{N}} \convol \tilde{E} /2$.
This means that the term itself is multipled by a factor of just
 $ \imath / 2\beta$, 
 and this transforming factor is what we need to use in the general case.

For a $\chi^{(2)}$ nonlinearity, 
 the time-domain displacement fields $D$ in the two polarizations are
~
\begin{eqnarray}
  D_x
&=&
  \epsilon_x \convol E_x 
 +
  2 \epsilon_0 \chi^{(2)} E_x E_y
,
\\
  D_y
&=&
  \epsilon_y \convol E_y
 +
  \epsilon_0 \chi^{(2)} E_x^2
.
\end{eqnarray}
Note in particular that the nonlinear part of the $D_y$ field
 does not depend on $E_y$, 
 making the wave eqn. (\ref{eqn-factored-2nd-nlRight-fwdbck})
 (aimed largely at a $\chi^{(3)}$ system)
 inappropriate.

In any case, 
 the $x$ nonlinear term is just $2 \epsilon_0 \chi^{(2)} E_x E_y$, 
 and the $y$ term $\epsilon_0 \chi^{(2)} E_x^2$
 so that in the pair of frequency domain wave equations
 (cf eqn. (\ref{eqn-2ndWave-simple})), 
 the nonlinear terms are 
~
\begin{eqnarray}
x: ~~~~ ~~~~
  \beta^2 \omega^2 \tilde{\mathscr{N}} \convol \tilde{E}
&\Leftrightarrow&
  2 \mu_0 \epsilon_0 \omega^2 \mathscr{F}\left[\chi^{(2)} E_x E_y\right]
\\
y: ~~~~ ~~~~
  \beta^2 \omega^2 \tilde{\mathscr{N}} \convol \tilde{E}
&\Leftrightarrow&
  \mu_0 \epsilon_0 \omega^2 \mathscr{F}\left[\chi^{(2)} E_x^2\right]
,
\end{eqnarray}
and in the factorised equations these become 
~
\begin{eqnarray}
x: ~~~~ ~~~~
    \imath \mu_0 \epsilon_0 
    \frac{\omega^2}{\beta(\omega)} 
    \mathscr{F}\left[\chi^{(2)} E_x E_y\right]
\\
y: ~~~~ ~~~~
    \imath \mu_0 \epsilon_0 
    \frac{\omega^2}{2 \beta(\omega)}  
    \mathscr{F}\left[\chi^{(2)} E_x^2\right]
.
\end{eqnarray}

Since we will want to apply the nonlinear effects in the time domain, 
 we need to back-transform these nonlinear terms:
~
\begin{eqnarray}
x: ~~~~ ~~~~
  \mathscr{F}^{-1}
  \left[
    \imath \mu_0 \epsilon_0 
    \frac{\omega^2}{\beta(\omega)} 
    \mathscr{F}\left[\chi^{(2)} E_x E_y\right]
  \right]
\\
y: ~~~~ ~~~~
  \mathscr{F}^{-1}
  \left[
  \imath \mu_0 \epsilon_0 
  \frac{\omega^2}{2 \beta(\omega)}  
  \mathscr{F}\left[\chi^{(2)} E_x^2\right]
  \right]
.
\end{eqnarray}

So we see that a true wideband approach to the nonlinearity 
 requires a pair of Fourier transforms.
In a semi-wideband limit around a central frequency $\omega_0$ 
 we can probably assume the factor $\omega^2/\beta(\omega)$ 
 becomes $c \omega/n(\omega_0)$. 
In an SVEA-like narrowband limit it would become
 $\omega_0^2/\beta(\omega_0) = c \omega_0 / n(\omega_0)$, 
 and the need for Fourier transforms vanishes.

If a more extensive collection of the $\chi^{(2)}$ coefficients 
 needs to be included, 
 this procedure can be reproduced
 using the appropriate nonlinear field conbinations.
If the time-response of the nonlinearity is also important, 
 then we can include this by replacing
 $\chi^{(2)} E_x E_y$ and $\chi^{(2)} E_x^2$ with appropriate convolutions:
 i.e. $(\chi^{(2)} \convol E_y ) E_x$ and $(\chi^{(2)} \convol E_x) E_x$.

%
\subsubsection{Factorization and envelopes}

Taking only the forward part of eqn (\ref{eqn-factored-2nd-nlRight-fwdbck}), 
 we replace $\tilde{E}_+(\omega) 
 = \tilde{A}_+(\omega+\omega_0) + \tilde{A}_+^*(\omega-\omega_0)$.
Since this the equation is linear in the derivatives, 
 when split into $\tilde{A}_+$ and $\tilde{A}_+^*$ parts
 it looks very similar, 
 being
~
\begin{eqnarray}
    \partial_z
    \tilde{A}_\pm
&=&
 \pm
    \imath
      \beta
    \tilde{A}_\pm
 \pm
  \frac{\imath \beta}{2}
  \tilde{\mathscr{N}}
  \convol
  \left[
    \tilde{A}_\pm
   +
    \tilde{A}_\mp
  \right]
.
\end{eqnarray}

For the case of a time-response $\chi^{(3)}$ nonlinearity, 
 the equation will be identical to that for the envelope 
 version of the
 directional $G^\pm$ fields, 
 i.e. eqn. (\ref{eqn-Ap-chi3}).

%
\subsubsection{Factorized fields}

An important feature of this approach is that we see that 
 \emph{any} contribution 
 (whether linear or not) 
 that is included in the source term
 will couple the forward and backward fields together.
Consider two differing factorisations
 of the same systems; 
 e.g. one with the loss included in $\beta$, 
 and one with it in the source term.
The one with the extra source contribution will see 
 a corresponding extra forward backward coupling term, 
 apparently conflicting with the fact that the two factorisations
 are of the same system.
The resolution of this conundrum is simply that the 
 forward and backward fields of the first factorisations ($E_{1\pm}$)
 are not the same as those for the second ($E_{2\pm}$); 
 the meaning of  ``forward field'' (or ``backward field'') 
 differs between the two implementations.
This is perhaps clearer in the $G^{\pm}$ formulation
 (see section \ref{S-directional}), 
 where the different factorisations would correspond to different 
 choices of the reference parameters $\alpha_r, \beta_r$.
If no further approximations have been made, 
 when the real electric and magnetic fields are reconstructed 
 from any factorised $E_{i\pm}$, 
 the answers should be in agreement.

%
\subsection{Transverse effects}
\label{Ss-secondorder-transverse}

In common with most pulse propagation, 
 we can restrict ourselves to paraxial beams and 
 incorporated transverse effects by 
 using a split step scheme alternating between the wave equation
 and the $\nabla_\perp^2$ diffraction term.
This is equally applicable to either the traditional or 
 factorization approaches.
However, 
 in the factorization approach we can treat the $\nabla_\perp^2 E$
 diffraction term as a ``source'' term, 
 and, 
 like the nonlinearity, 
 move it to the RHS before factorising.
Thus eqn. (\ref{eqn-factored-2nd-nlRight-fwdbck}) could be
 rewritten to include diffraction as
~
\begin{eqnarray}
    \partial_z
    \tilde{E}_\pm
&=&
 \pm
    \imath
      \beta
    \tilde{E}_\pm
 \pm
  \frac{\imath \beta}{2}
  \tilde{\mathscr{N}}
  \convol
  \left[
    \tilde{E}_+ + \tilde{E}_-
  \right]
\nonumber
\\
&&
~~~~
~~~~
~~~~
~~~~
 \pm
  \frac{\imath}{2\beta}
  \nabla_\perp^2
  \left[
    \tilde{E}_+ + \tilde{E}_-
  \right]
.
\label{eqn-factored-2nd-nlRight-fwdbck-diff}
\end{eqnarray}

%
\section{Forward-backward coupling}\label{S-forwardbackward}

We can see in 
 eqns.(\ref{eqn-firstorder-Gpm-comoving},
       \ref{eqn-firstorder-Epm-comoving},
       \ref{eqn-factored-2nd-nlRight-fwdbck})
 that we simplify into a forward-only picture by dropping the part of the 
 nonlinear polarization term due to the backward field.
In situations where there is no pre-existing backward field, 
 and where there are no interfaces to cause reflection, 
 this is an excellent approximation that holds true
 in the regime of weak nonlinearity.
It is only an approximation, 
 because the nonlinear polarization drives {\em both} the forward 
 and backward fields, 
 so in strongly nonlinear systems, 
 a backward wave can be generated directly by the forward wave.
The important ``weak nonlinearity'' criteria for perturbative nonlinearities 
 to guarantee the validity of a forward-only model is \cite{Kinsler-2007josab}
~
\begin{eqnarray}
  \frac{1}{n_0^2}
  \sum_{m>1} m \chi^{(m)} E^{m-1} 
&\ll&
  1.
\end{eqnarray}

On the subject of reflections from interfaces, 
 it is worth noting that ``nonlinear'' reflections can occur
 even if the linear dispersion on both sides is identical --
 as long as the nonlinearity changes, 
 as in e.g. periodic poling, 
 where its sign changes.

Note that since nonlinearities are in practice very weak 
 (e.g. $\chi^{(3)} E^{3} \sim 0.06$ at the damage threshold of fused silica), 
 uni-directional propagation models perform very well, 
 and the role of nonlinear reflections is generally negligible.

%
\section{Conclusions}\label{S-conclusion}

I have described three forms for the spatial propagation of 
 optical fields: Maxwell's equations, directional fields, 
 and second order wave equations.
These forms have been describe in both standard and envelope-carrier
 pictures.
While solving Maxwell's equations remains the ``gold standard'' 
 and most exact procedure, 
 it is computationally demanding, 
 and it can be difficult to set up initial conditions.
These difficulties are avoided by using a directional fields 
 approach, 
 where we can propagate more efficiently in the usual 
 forward-only cases.
Further, 
 envelope theories based on forward-only directional fields 
 give equations of motion similar in form to the 
 traditional SVEA ones based on the second order wave equation, 
 but without requiring complicated approximations.

When comparing the various approaches taken to directional fields, 
 a number of important points stand out.

\begin{enumerate}

\item 
 The first successful attempt at deriving useful directional 
 versions of Maxwell's equations 
 was by Kolesik et al. \cite{Kolesik-MM-2002prl,Kolesik-M-2004pre}.

\item
 The most flexible and complete formulation is 
 the directional $G^\pm$ fields
 of Kinsler et al. \cite{Kinsler-2010pra-dblnlGpm,Kinsler-RN-2005pra}, 
 relying only on simple combinations of Maxwell's equations
 to achieve a directional form.
It is applicable to propagation media
 with {\em any} frequency-dependent electric or magnetic properties, 
 and variant forms \cite{Kinsler-2006arXiv-fleck} can be used if required.

\item 
 The factorization style approach 
 \cite{Blow-W-1989jqe,Ferrando-ZCBM-2005pre,Genty-KKD-2007oe,Kinsler-2010pra-fchhg} 
 gives propagation equations for
 the electric field that can be simply expressed and solved 
 without the construction
 of the conceptually abstract $G^\pm$ directed fields, 
 even for media with a magnetic response \cite{Kinsler-2010pra-fchhg}.

\end{enumerate}

It is encouraging that these three approaches discussed in this paper 
 (Maxwell's equations, 
 directional $G^\pm$ fields, 
 and factorized second order wave equations) 
 all give essentially 
 identical results in the case of uni-directional propagation in 
 non-magnetic media.

%
\section{Acknowledgments}

I acknowledge a wide variety of useful 
 discussions with G.H.C. New, S.B.P. Radnor, J.M. Dudley, 
 and G. Genty.
I also thank N. Broderick for bringing 
 refs. \cite{Sipe-PD-1994josaa,deSterke-SS-1996pre}
 to my attention; 
 and to M. Scalora for 
 refs. \cite{Dowling-SBB-1994jap,Scalora-DBB-1994prl,Scalora-C-1994oc}.

%

\end{document}